\documentclass[12pt,letterpaper]{article}
\usepackage{amsmath,amssymb,array,calc,rotating,epsfig,psfrag,amscd, cite}

\makeatletter
\renewcommand\section{\@startsection {section}{1}{\z@}%
                                   {-3.5ex \@plus -1ex \@minus -.2ex}
                                   {2.3ex \@plus.2ex}%
                                   {\normalfont\large\bfseries}}
\renewcommand\subsection{\@startsection{subsection}{2}{\z@}%
                                     {-3.25ex\@plus -1ex \@minus -.2ex}%
                                     {1.5ex \@plus .2ex}%
                                     {\normalfont\bfseries}}
\makeatother

\let\non\nonumber

\let\S=\Sigma

\newcommand{\bea}{\begin{eqnarray}}
\newcommand{\eea}{\end{eqnarray}}
\newcommand{\be}{\begin{equation}}
\newcommand{\ee}{\end{equation}}

\newcommand{\m}{\mu}

\newcommand{\p}{\partial}

\newcommand{\C}[1]{$(\ref{#1})$}

\def\IZ{\relax\ifmmode\mathchoice
{\hbox{\cmss Z\kern-.4em Z}}{\hbox{\cmss Z\kern-.4em Z}}
{\lower.9pt\hbox{\cmsss Z\kern-.4em Z}} {\lower1.2pt\hbox{\cmsss
Z\kern-.4em Z}}\else{\cmss Z\kern-.4em Z}\fi}
\def\IR{\relax{\rm I\kern-.18em R}}

\def\one{{\hbox{ 1\kern-.8mm l}}}

\newlength{\bredde}
\def\slash#1{\settowidth{\bredde}{$#1$}\ifmmode\,\raisebox{.15ex}{/}
\hspace*{-\bredde} #1\else$\,\raisebox{.15ex}{/}\hspace*{-\bredde}
#1$\fi}

\newsavebox{\zzzbar}
\sbox{\zzzbar}
  {\setlength{\unitlength}{0.9em}
  \begin{picture}(0.6,0.7)
  \thinlines
  \put(0,0){\line(1,0){0.6}}
  \put(0,0.75){\line(1,0){0.575}}
  \multiput(0,0)(0.0125,0.025){30}{\rule{0.3pt}{0.3pt}}
  \multiput(0.2,0)(0.0125,0.025){30}{\rule{0.3pt}{0.3pt}}
  \put(0,0.75){\line(0,-1){0.15}}
  \put(0.015,0.75){\line(0,-1){0.1}}
  \put(0.03,0.75){\line(0,-1){0.075}}
  \put(0.045,0.75){\line(0,-1){0.05}}
  \put(0.05,0.75){\line(0,-1){0.025}}
  \put(0.6,0){\line(0,1){0.15}}
  \put(0.585,0){\line(0,1){0.1}}
  \put(0.57,0){\line(0,1){0.075}}
  \put(0.555,0){\line(0,1){0.05}}
  \put(0.55,0){\line(0,1){0.025}}
  \end{picture}}

\newcommand{\ena}{\end{eqnarray}}
\newcommand{\beqa}{\begin{eqnarray}}
\newcommand{\eeqa}{\end{eqnarray}}

\def\m{\mu}

\def\S{\Sigma}

\begin{document}
\begin{titlepage}

\begin{center}



\vskip 2 cm
{\Large \bf Proving relations between modular graph functions }\\
\vskip 1.25 cm { Anirban Basu\footnote{email address:
    anirbanbasu@hri.res.in} } \\
{\vskip 0.5cm Harish--Chandra Research Institute, Chhatnag Road, Jhusi,\\
Allahabad 211019, India\\}

\end{center}

\vskip 2 cm

\begin{abstract}
\baselineskip=18pt

We consider modular graph functions that arise in the low energy expansion of the four graviton amplitude in type II string theory. The vertices of these graphs are the positions of insertions of vertex operators on the toroidal worldsheet, while the links are the scalar Green functions connecting the vertices.
Graphs with four and five links satisfy several non--trivial relations, which have been proved recently.  We prove these relations by using elementary properties of Green functions and the details of the graphs. We also prove a relation between modular graph functions with six links.   

\end{abstract}

\end{titlepage}


\section{Introduction}

Calculating amplitudes in perturbative string theory is an important tool to analyze terms in the effective action of string theory. At a fixed order in the genus expansion, they yield local and non--local terms at all orders in the $\alpha'$ expansion. Knowledge of these amplitudes also plays a significant role in understanding non--perturbative duality symmetries of string theory because duality covariant couplings of terms in the effective action must reproduce these amplitudes when expanded around weak coupling. While amplitudes at tree level have been obtained for several processes, amplitudes at higher genus have not been so well studied. Here we consider certain local terms in the low energy expansion of the four graviton amplitude at genus one in type II string theory in ten dimensions. The low energy expansion yields terms of the form $D^{2k} \mathcal{R}^4$, where $\mathcal{R}^4$ represents a specific contraction of four powers of the Weyl tensor and $D^{2k}$ represents $2k$ derivatives. For fixed $k$, the evaluation of the genus one amplitude amounts to evaluating integrals of the form
\be \sum_i \int_{\mathcal{F}_L} \frac{d^2\tau}{\tau_2^2} f_{k,i}(\tau,\bar\tau)\ee     
where $\tau$ is the complex structure modulus of the torus, and $d^2\tau = d\tau_1 d\tau_2$. We have integrated over the truncated fundamental domain of $SL(2,\mathbb{Z})$ defined by~\cite{Green:1999pv}
\be \label{one}\mathcal{F}_L = \Big\{ -\frac{1}{2} \leq \tau_1 \leq \frac{1}{2}, \vert \tau \vert \geq 1, \tau_2 \leq L\Big\},\ee  
where $L \rightarrow \infty$, which produces finite as well as contributions that diverge as $L\rightarrow \infty$. The finite contributions are the required local contributions while the divergent contributions cancel those from the boundary of moduli space which have to be calculated separately. In \C{one}, each $f_{k,i}(\tau,\bar\tau)$ is a nonholomorphic modular form that is $SL(2,\mathbb{Z})$ invariant, and is referred to as a modular graph function. The sum over $i$ runs over a finite number of terms that is determined by the various graphs that arise at that order in $k$. The vertices of these graphs are the positions of insertions of the vertex operators on the toroidal worldsheet, while the links are the scalar Green functions that connect the various vertices\footnote{The modular graph functions we consider and the relations we derive between them do not involve derivatives of scalar Green functions. They show up, for example, in the low energy expansion of the five graviton amplitude~\cite{Green:2013bza}.}. It is important to have a detailed understanding of the modular graph functions in order to obtain the genus one amplitudes.   

While these modular graph functions at leading orders in the low energy expansion are not difficult to evaluate, they become quite involved at higher orders in the momentum expansion. We shall consider graphs at orders $D^8\mathcal{R}^4$ and $D^{10} \mathcal{R}^4$, and also one at order $D^{12}\mathcal{R}^4$, that arise in the derivative expansion. Graphs at order $D^{2k} \mathcal{R}^4$ have $k$ links that arise from the $k$ scalar Green functions from the Koba--Nielsen factor in the low energy expansion. For the cases we consider, it turns out that they satisfy various non--trivial relations at each order in the derivative expansion. These relations were originally conjectured based on Poisson equations the modular graph functions satisfy, and their asymptotic expansions~\cite{D'Hoker:2015foa}. Subsequently they have been proven using various techniques based on intricate details of their modular properties~\cite{D'Hoker:2015zfa,D'Hoker:2016jac}. These relations should also follow from identities between elliptic polylogarithms~\cite{D'Hoker:2015qmf}. Importantly, these relations are between graphs having the same number of links, though not necessarily the same number of vertices. 
      
 The principal idea behind these relations between the modular graph functions is that each $f_{k,i}$ satisfies a Poisson equation. These Poisson equations for the graphs with cubic vertices and six links for terms that are relevant for the $D^{12}\mathcal{R}^4$ interaction have been derived in~\cite{Basu:2015ayg,Basu:2016xrt}. These are the Mercedes diagram and the three loop ladder diagram\footnote{The Poisson equation for the three loop ladder diagram has a source term that has a modular graph function with two derivatives, which arises in the five graviton amplitude.}. In obtaining these Poisson equations, diagrammatic rather then algebraic expressions for the various graphs have been the starting point. The Poisson equations have been obtained by manipulating them using various properties of the Green functions. Thus this line of analysis is very different from the ones that have been followed in the papers mentioned above. In this work, we generalize this approach to derive the Poisson equations for all the modular graph functions that are relevant for the $D^8 \mathcal{R}^4$ and $D^{10} \mathcal{R}^4$ interactions. These then lead to relations between graphs with four links and similarly with five links, providing an alternate proof of these relations. We then consider the Poisson equation for the Mercedes graph, and derive Poisson equations for certain other graphs with five vertices and six links. They lead to a relation between modular graph functions with six links. Our analysis should be generalizable to interactions at higher orders in the derivative expansion as well, and hence should provide more relations between modular graph functions.   

We start with a very brief review of the genus one four graviton amplitude in type II string theory in ten dimensions, and the various properties of the scalar Green function we shall need. We then derive the relations between the modular graph functions with four, five and six links.    

\section{The four graviton amplitude and the scalar Green function}

The local terms arise from the low momentum expansion of the four graviton amplitude at genus one in type II superstring theory in ten dimensions given by
\be \mathcal{A}_4 = 2\pi \mathcal{I}(s,t,u) \mathcal{R}^4, \ee
where
\be \label{oneloop}\mathcal{I} (s,t,u) = \int_{\mathcal{F}_L} \frac{d^2\tau}{\tau_2^2} F(s,t,u;\tau,\bar\tau),\ee
where the Mandelstam variables $s, t, u$ satisfy the on--shell relation
\be s+t+u=0.\ee 
The factor $F(s,t,u;\tau,\bar\tau)$ which encodes the worldsheet moduli and momentum dependence is given by
\be \label{D}F(s,t,u;\tau,\bar\tau) = \prod_{i=1}^4 \int_\S  \frac{d^2 z^{(i)}}{\tau_2} e^{\mathcal{D}},\ee
where $z^{(i)}$ $(i=1,2,3,4)$ are the positions of insertions of the four vertex operators on the toroidal worldsheet $\S$. Thus $d^2 z^{(i)} = d({\rm Re} z^{(i)}) d({\rm Im}z^{(i)})$, where
\be -\frac{1}{2} \leq {\rm Re} z^{(i)} \leq \frac{1}{2}, \quad 0 \leq {\rm Im} z^{(i)}\leq \tau_2 \ee
for all $i$. In \C{D}, the expression for $\mathcal{D}$ is given by
\be \label{defD}4\mathcal{D} = \alpha' s ({G}_{12} + {G}_{34})+\alpha' t ({G}_{14} + {G}_{23})+ \alpha' u ({G}_{13} +{G}_{24}),\ee
where ${G}_{ij}$ is the scalar Green function on the torus with complex structure $\tau$ between points $z^{(i)}$ and $z^{(j)}$ after removing an irrelevant zero mode contribution.
Its explicit expression is given by~\cite{Green:1999pv,Green:2008uj}
\be \label{Green}G(z;\tau) = \frac{1}{\pi} \sum_{(m,n)\neq(0,0)} \frac{\tau_2}{\vert m\tau+n\vert^2} e^{\pi[\bar{z}(m\tau+n)-z(m\bar\tau+n)]/\tau_2},\ee
where $G(z;\tau)$ is modular invariant, and single valued. Thus
\be \label{sv}G(z;\tau) = G(z+1;\tau) = G(z+\tau;\tau).\ee

Thus the local terms are obtained by expanding the exponential involving the Green functions and performing the various integrals. Contributions upto the $D^{10} \mathcal{R}^4$ interaction have been obtained in~\cite{D'Hoker:2015foa,Basu:2016fpd} in ten dimensions as well as for toroidal compactifications. For the later case, the amplitudes satisfy Poisson equations with respect to the spacetime moduli which has to be solved with appropriate boundary conditions.   

In order to derive the Poisson equations, we make use of the various properties satisfied by the Green function $G_{ij}$ (see~\cite{D'Hoker:2015foa,Basu:2015ayg} for various details). We find it very useful to use the relations satisfied by them under the variation of the Beltrami differential $\m$. We have that
\be \label{onevar} \p_\mu G(z_1,z_2) = -\frac{1}{\pi} \int_\S d^2 z \p_z G(z,z_1) \p_z G(z,z_2),\ee  
and
\be \label{novar}\bar\p_{\m}\p_\m G(z_1,z_2) =0.\ee
Also the $SL(2,\mathbb{Z})$ invariant Laplacian is defined by
\be \label{beltrami}\Delta = 4\tau_2^2\frac{\p^2}{\p\tau\p\bar\tau} = \bar\p_{\m} \p_\m.\ee
The Green function satisfies the equations
\bea \label{eigen}\bar{\p}_w\p_z G(z,w) = \pi \delta^2 (z-w) - \frac{\pi}{\tau_2}, \non \\ \bar{\p}_z\p_z G(z,w) = -\pi \delta^2 (z-w) + \frac{\pi}{\tau_2} \eea
which is repeatedly used in our analysis. 

In the various manipulations, we often obtain expressions involving $\p_z G(z,w)$ where $z$ is integrated over $\S$. We then integrate by parts without picking up boundary contributions on $\S$ as $G(z,w)$ is single valued. Hence we also drop all contributions which are total derivatives as they vanish. Also we readily use $\p_z G(z,w) = -\p_w G(z,w)$ using the translational invariance of the Green function. Finally, we have that
\be \int_\S d^2 z G(z,w)=0\ee
which easily follows from \C{Green}.

In the various expressions, for brevity we write
\be \int_\S d^2 z \int_\S d^2 w \ldots \equiv \int_{zw\ldots}.\ee
We shall also drop contributions that vanish due to simple manipulations.

We shall find it very useful to denote the various graphs diagrammatically. In these graphs, the notations for holomorphic and antiholomorphic derivatives with respect to the worldsheet coordinate acting on the Green function are given in figure 1. From the structure of \C{Green} it follows that one particle reducible diagrams vanish and hence we ignore them. 

\begin{figure}[ht]
\begin{center}
\[
\mbox{\begin{picture}(240,60)(0,0)
\includegraphics[scale=.75]{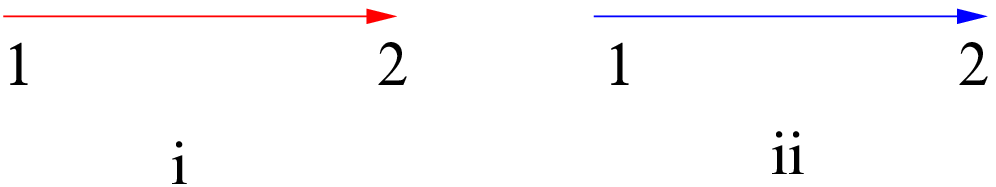}
\end{picture}}
\]
\caption{(i) $\p_2 G_{12} = -\p_1 G_{12}$, (ii) $\bar\p_2 G_{12} = -\bar\p_1 G_{12}$}
\end{center}
\end{figure}

In the various diagrams, $\mu$ along a link stands for $\p_\mu$, while $\bar\mu$ stands for $\bar\p_{\mu}$. We shall follow the conventions of~\cite{Green:2008uj} in naming the various modular graph functions.

\section{The elementary diagrams}

Some of the diagrams that are relevant to our analysis can be calculated very easily and for those we simply give the final answers. These diagrams  are given in figure 2.

\begin{figure}[ht]
\begin{center}
\[
\mbox{\begin{picture}(360,90)(0,0)
\includegraphics[scale=.95]{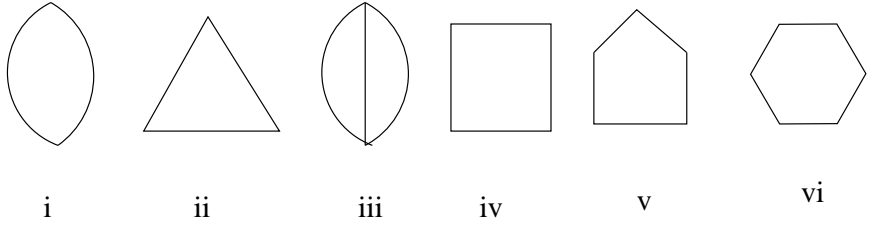}
\end{picture}}
\]
\caption{The diagrams (i) $D_2$, (ii) $D_{1,1,1}$, (iii) $D_3$, (iv) $D_{1,1,1,1}$, (v) $D_{1,1,1,1,1}$ and (vi) $D_{1,1,1,1,1,1}$}
\end{center}
\end{figure}

The last three diagrams $D_{1,1,1,1}$, $D_{1,1,1,1,1}$ and $D_{1,1,1,1,1,1}$ first arise at orders $D^8\mathcal{R}^4$, $D^{8} \mathcal{R}^5$ and $D^{8}\mathcal{R}^6$ respectively.

Now these diagrams are defined by
\bea && D_2 = \frac{1}{\tau_2^2} \int_{12} G_{12}^2, \quad D_{1,1,1} = \frac{1}{\tau_2^3} \int_{123} G_{12} G_{23} G_{13}, \quad D_3= \frac{1}{\tau_2^2} \int_{12} G_{12}^3, \non \\ && D_{1,1,1,1} = \frac{1}{\tau_2^4} \int_{1234} G_{12} G_{23} G_{34} G_{14}, \quad D_{1,1,1,1,1} = \frac{1}{\tau_2^5} \int_{12345} G_{12} G_{23} G_{34} G_{45} G_{15}, \non \\ && D_{1,1,1,1,1,1} = \frac{1}{\tau_2^6}\int_{123456} G_{12} G_{23} G_{34} G_{45} G_{56} G_{16}.\eea

It is easy to determine the equations for $D_2$, $D_{1,1,1}$, $D_3$, $D_{1,1,1,1}$, $D_{1,1,1,1,1}$ and $D_{1,1,1,1,1,1}$. Using the equations for the variations of the Green function under change of Beltrami differentials mentioned above, we get that
\bea && \Delta D_2 = 2 D_2, \quad \Delta D_{1,1,1} = 6 D_{1,1,1}, \quad \Delta D_3 = 6 D_{1,1,1}, \non \\  && \Delta D_{1,1,1,1} = 12 D_{1,1,1,1}, \quad \Delta D_{1,1,1,1,1} = 20 D_{1,1,1,1,1}, \quad \Delta D_{1,1,1,1,1,1} = 30 D_{1,1,1,1,1,1}.\non \\ \eea
This leads to the solutions
\bea \label{1} && D_2 = E_2, \quad D_{1,1,1} = E_3, \quad D_3 = E_3 + \zeta(3), \non \\ && D_{1,1,1,1} = E_4, \quad D_{1,1,1,1,1} = E_5, \quad D_{1,1,1,1,1,1} = E_6,\eea
based on boundary conditions and the asymptotic expansions at large $\tau_2$~\footnote{The relevant asymptotic expansions of the various graph functions are given in~\cite{Green:2008uj,D'Hoker:2015foa}.}. Here $E_s$ is the non--holomorphic Eisenstein series defined by
\be E_s (\tau,\bar\tau) = \sum_{(m,n)\neq (0,0)}\frac{\tau_2^s}{\pi^s\vert m+n\tau\vert^{2s}}.\ee        

\section{Poisson equations for diagrams with four links}

We first consider the Poisson equations for non--trivial diagrams with four links. These are diagrams that are relevant at order $D^8\mathcal{R}^4$ in the low momentum expansion, and are given in figure 3. 
\begin{figure}[ht]
\begin{center}
\[
\mbox{\begin{picture}(120,90)(0,0)
\includegraphics[scale=.95]{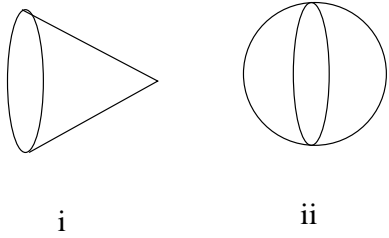}
\end{picture}}
\]
\caption{The diagrams (i) $D_{1,1,2}$ and (ii) $D_4$}
\end{center}
\end{figure}

Thus we have that
\bea D_{1,1,2} = \frac{1}{\tau_2^3}\int_{123} G_{12}^2 G_{13} G_{23}, \quad D_4 = \frac{1}{\tau_2^2} \int_{12} G_{12}^4.\eea
In fact, $D_{1,1,2}$ does not appear in the expression for the $D^8\mathcal{R}^4$ term, however it does appear in the final relation involving the modular graph functions.  

\subsection{The Poisson equation for $D_{1,1,2}$}

We first obtain the Poisson equation for $D_{1,1,2}$. 
From \C{novar} and \C{beltrami} have that
\be \Delta D_{1,1,2} = 2F_1 + 2 F_2 + 4 (F_3 + c.c.),\ee
where
\bea F_1 &=& \frac{1}{\tau_2^3} \int_{123} \p_\m G_{12} \bar\p_\mu G_{12} G_{13}G_{23}, \non \\ F_2 &=& \frac{1}{\tau_2^3} \int_{123} G_{12}^2 \p_\m G_{13} \bar\p_\m G_{23}, \non \\ F_3 &=& \frac{1}{\tau_2^3} \int_{123} G_{12} \p_\m G_{12}\bar\p_\m G_{13} G_{23}\eea
and are given in figure 4.

\begin{figure}[ht]
\begin{center}
\[
\mbox{\begin{picture}(280,90)(0,0)
\includegraphics[scale=.9]{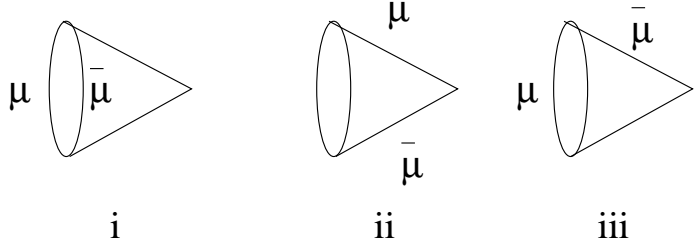}
\end{picture}}
\]
\caption{The diagrams (i) $F_1$, (ii) $F_2$ and (iii) $F_3$}
\end{center}
\end{figure}

We now manipulate them using \C{onevar} and \C{eigen} to express them in terms of various modular graph functions. This leads to
\bea F_1 &=& \frac{3}{2} E_2^2 -\frac{3}{2} E_4- 2 D_{1,1,2}, \non \\ F_2 &=& D_{1,1,2}, \non \\ F_3 &=& -\frac{1}{2} E_2^2 + \frac{3}{2} E_4 +\frac{1}{2}D_{1,1,2}.\eea
Adding the various contributions we get that
\be \label{e1}(\Delta -2) D_{1,1,2} = 9 E_4 - E_2^2\ee
which has been deduced in~\cite{D'Hoker:2015foa} using different techniques. 

\subsection{The Poisson equation for $D_4$}

We next obtain the Poisson equation for $D_4$. It is not particularly useful to start directly with the diagram as given in figure 3 and analyze variations of the Green functions in the diagram. Hence we proceed differently. 

We start with the diagram $F_4$ which is given by
\be F_4 = \frac{1}{\tau_2^2} \int_{123} \bar\p_1 \p_2 G_{12} G_{13} \p_\m G_{13} G_{23} \bar\p_\mu G_{23}\ee
as shown in figure 5.

\begin{figure}[ht]
\begin{center}
\[
\mbox{\begin{picture}(100,110)(0,0)
\includegraphics[scale=.5]{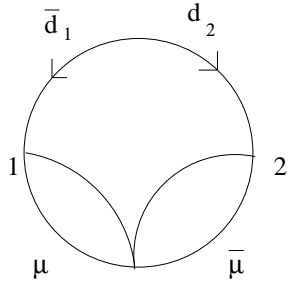}
\end{picture}}
\]
\caption{The diagram $F_4$}
\end{center}
\end{figure}

This can be evaluated by simply using the relation for the Green function \C{eigen} for the $\p$ and $\bar\p$ on the same link leading to
\be F_4 = \pi F_5 - \pi F_6 F_6^*,\ee
where the diagrams $F_5$ and $F_6$ are given in figure 6. While $F_5$ is obtained from the variation of $D_4$, $F_6$ is obtained from the variation of $D_2$. 

Now $F_4$ can also be evaluated by moving the $\p$ and the $\bar\p$ along the links appropriately leading to
\be \frac{1}{\pi} F_4 = -\frac{1}{12} D_4 -\frac{5}{4} E_2^2 + 2E_4 + D_{1,1,2} + \frac{1}{\pi^2}F_7,\ee
where $F_7$ is given by
\be F_7 = \frac{1}{\tau_2^2} \int_{1234} \p_1 G_{12} \bar\p_1 G_{12}\p_1 G_{13}  \bar\p_1 G_{14} G_{23} G_{24}  \ee 
as depicted in figure 7.

\begin{figure}[ht]
\begin{center}
\[
\mbox{\begin{picture}(180,120)(0,0)
\includegraphics[scale=.75]{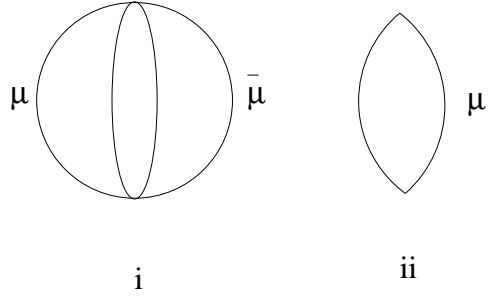}
\end{picture}}
\]
\caption{The diagrams (i) $F_5$ and (ii) $F_6$}
\end{center}
\end{figure}

\begin{figure}[ht]
\begin{center}
\[
\mbox{\begin{picture}(70,70)(0,0)
\includegraphics[scale=.9]{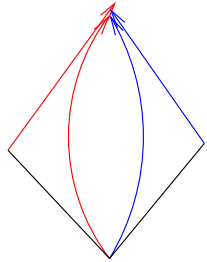}
\end{picture}}
\]
\caption{The diagram $F_7$}
\end{center}
\end{figure}

We now evaluate the diagram $F_7$. However rather than do so directly, we find it convenient to start from the diagram $F_8$ which is given by
\be F_8= \frac{1}{\tau_2^2}\int_{12345}\bar\p_1 \p_2 G_{12} \p_1 G_{15} \bar\p_2 G_{25} \p_1 G_{13} \bar\p_2 G_{24} G_{35} G_{45} \ee
as shown in figure 8.

\begin{figure}[ht]
\begin{center}
\[
\mbox{\begin{picture}(90,110)(0,0)
\includegraphics[scale=.65]{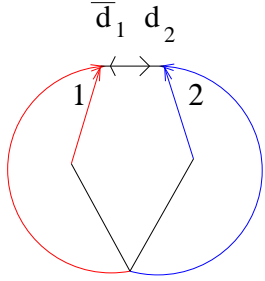}
\end{picture}}
\]
\caption{The diagram $F_8$}
\end{center}
\end{figure}

As before, we can evaluate it in two ways. Using \C{eigen} for the $\p$ and $\bar\p$ on the same link trivially leads to
\be F_8 = \pi F_7 - \pi^3 F_6 F_6^* .\ee 
On the other hand, evaluating it by moving the derivatives through the links leads to
\be \frac{1}{\pi^3} F_8 =   - D_{1,1,2} + E_4 +\frac{1}{4} D_4 - \frac{1}{4} E_2^2.\ee
Hence substituting the various relations, this gives us 
\be \label{z1}F_5 - 2 F_6 F_6^* = \frac{1}{6} D_4 + 3 E_4 - \frac{3}{2} E_2^2.\ee
Now we can obtain the Poisson equation involving $D_4$. We note that 
\be \label{Y}\Delta (D_4 - 3 E_2^2) = \p_\m \bar\p_\m (D_4 - 3 E_2^2) = 12(F_5 - 2 F_6 F_6^*) -12 E_2^2.\ee
In \C{Y} we have chosen the relative factors of $D_4$ and $E_2^2$ appropriately such that $\Delta$ acting on that combination yields $F_5- 2 F_6 F_6^*$ which arises in our analysis. Thus from \C{z1}
we get the Poisson equation
\be \label{e2}(\Delta -2)(D_4 - 3 E_2^2) = 36 E_4 - 24 E_2^2\ee
as conjectured in~\cite{D'Hoker:2015foa}.

Thus from \C{e1}, \C{e2} and the asymptotic expansions, we get the relation 
\be D_4 = 24 D_{1,1,2} + 3 E_2^2 - 18 E_4\ee
between various modular graph functions with four links as conjectured in~\cite{D'Hoker:2015foa}.

This strategy used in obtaining the Poisson equations will be used repeatedly in our analysis. We shall often have to manipulate diagrams whose variations using \C{onevar} do not lead to particularly useful expressions. We shall instead manipulate appropriately chosen auxiliary diagrams involving more links and derivatives which reduce to the parent diagrams trivially using \C{eigen}. These auxiliary diagrams are then evaluated independently such that they are expressible in terms of modular graph functions involving no derivatives at all. This helps us in achieving considerable simplification in obtaining the Poisson equations for the various diagrams.

\section{Poisson equations for diagrams with five links}

We next consider Poisson equations for non--trivial diagrams with five links. They arise at order $D^{10}\mathcal{R}^4$ in the low momentum expansion. The relevant diagrams with five links are given in figure 9.
 
\begin{figure}[ht]
\begin{center}
\[
\mbox{\begin{picture}(180,160)(0,0)
\includegraphics[scale=.8]{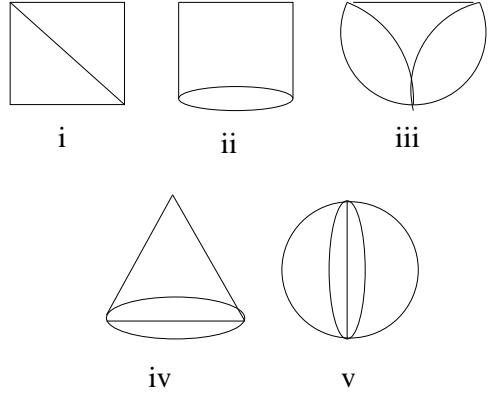}
\end{picture}}
\]
\caption{The diagrams (i) $D_{1,1,1,1;1}$, (ii) $D_{1,1,1,2}$, (iii) $D_{1,2,2}$, (iv) $D_{1,1,3}$ and (v) $D_5$}
\end{center}
\end{figure}

Thus they are given by the expressions
\bea && D_{1,1,1,1;1} = \frac{1}{\tau_2^4} \int_{1234} G_{12} G_{23} G_{34} G_{14} G_{13}, \quad D_{1,1,1,2} = \frac{1}{\tau_2^4} \int_{1234} G_{12} G_{23} G_{34}^2 G_{14}, \non \\ && D_{1,2,2} = \frac{1}{\tau_2^3}\int_{123} G_{12} G_{23}^2 G^2_{13}, \quad D_{1,1,3} = \frac{1}{\tau_2^3}\int_{123} G_{12} G_{13} G_{23}^3, \non \\ && D_5 = \frac{1}{\tau_2^2}\int_{12} G_{12}^5.\eea
We now obtain the Poisson equations for each of these diagrams.

\subsection{The Poisson equation for $D_{1,1,1,1;1}$}

We first obtain the Poisson equation for $D_{1,1,1,1;1}$. Proceeding as before, we have that
\be \Delta D_{1,1,1,1;1} = 4 (F_9 + c.c.) + 4(F_{10} + 2 F_{11})\ee
where
\bea F_9 &=& \frac{1}{\tau_2^4} \int_{1234} \p_\m G_{12} G_{23} G_{34} G_{14} \bar\p_\m G_{13}, \non \\ F_{10} &=& \frac{1}{\tau_2^4} \int_{1234} \p_\m G_{12} \bar\p_\m G_{23} G_{34} G_{14} G_{13}, \non \\ F_{11} &=& \frac{1}{\tau_2^4} \int_{1234}  \p_\m G_{12} G_{23} G_{34} \bar\p_\m G_{14} G_{13} \eea
which is given in figure 10. 

\begin{figure}[ht]
\begin{center}
\[
\mbox{\begin{picture}(260,130)(0,0)
\includegraphics[scale=.7]{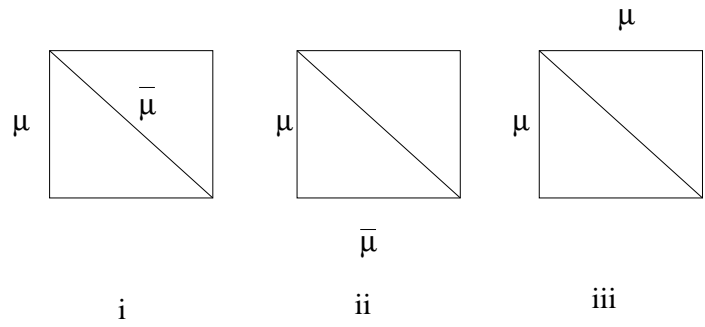}
\end{picture}}
\]
\caption{The diagrams (i) $F_9$, (ii) $F_{10}$ and (iii) $F_{11}$}
\end{center}
\end{figure}

Each of these diagrams can be manipulated to give us expressions involving modular graph functions with no derivatives. Using \C{onevar} and \C{eigen} we get that
\bea F_9 +c.c. &=& -D_{1,1,1,1;1} -2D_{1,1,1,2} + 2 E_2 E_3 - 2 E_5, \non \\ F_{10} &=& D_{1,1,1,1;1}, \non \\ F_{11} &=& 2 E_5 + D_{1,1,1,2} - E_2 E_3.\eea

Thus we get the Laplace equation
\be \Delta D_{1,1,1,1;1} = 8 E_5, \ee
leading to
\be \label{R1}D_{1,1,1,1;1} = \frac{2}{5}E_5 +\frac{\zeta(5)}{30}\ee
on using the asymptotic expansion, as derived in~\cite{D'Hoker:2015foa} using other techniques. 

\subsection{The Poisson equation for $D_{1,1,1,2}$}

We next obtain the Poisson equation for $D_{1,1,1,2}$. Proceeding as before, we have that
\be \Delta D_{1,1,1,2} = 2[F_{12} + 3(F_{13} + c.c.) + 3F_{14} ],\ee
where
\bea F_{12} &=& \frac{1}{\tau_2^4} \int_{1234} G_{12} \p_\m G_{23} \bar\p_\m G_{23} G_{34} G_{14}, \non \\ F_{13} &=& \frac{1}{\tau_2^4} \int_{1234} G_{12} G_{23} \p_\m G_{23} G_{34} \bar\p_\m G_{14}, \non \\ F_{14} &=& \frac{1}{\tau_2^4} \int_{1234} \p_\m G_{12} G_{23}^2 \bar\p_\m G_{34} G_{14} \eea
as given in figure 11. 

\begin{figure}[ht]
\begin{center}
\[
\mbox{\begin{picture}(220,90)(0,0)
\includegraphics[scale=.8]{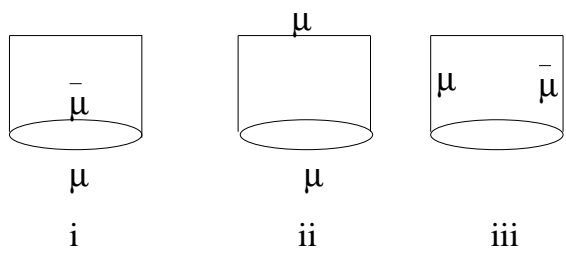}
\end{picture}}
\]
\caption{The diagrams (i) $F_{12}$, (ii) $F_{13}$ and (iii) $F_{14}$}
\end{center}
\end{figure}

Manipulating these diagrams we get that
\bea F_{12} &=& -\frac{3}{2} D_{1,1,1,1;1} + E_2 E_3 - E_5, \non \\ F_{13}+c.c. &=& 3 E_5 + D_{1,1,1,1;1} -E_2 E_3, \non \\ F_{14} &=& D_{1,1,1,2} .\eea

Thus we obtain the Poisson equation
\bea \label{R2}(\Delta-6) D_{1,1,1,2} &=& 3 D_{1,1,1,1;1} - 4 E_2 E_3 + 16 E_5 \non \\ &=& \frac{86}{5} E_5 - 4 E_2 E_3 +\frac{\zeta(5)}{10}\eea
as derived in~\cite{D'Hoker:2015foa} using different techniques. 

\subsection{The Poisson equation for $D_{1,2,2}$}

We next obtain the Poisson equation for $D_{1,2,2}$. Like the earlier analysis, we have that
\be \Delta D_{1,2,2} = 4[F_{15} + (F_{16} +c.c.) + 2 F_{17}]\ee
where
\bea F_{15} &=& \frac{1}{\tau_2^3}\int_{123} \p_\m G_{12} \bar\p_\m G_{12} G_{23}^2 G_{13}, \non \\ F_{16} &=& \frac{1}{\tau_2^3}\int_{123} G_{12} \p_\m G_{12} G_{23}^2 \bar\p_\m G_{13}, \non \\ F_{17} &=& \frac{1}{\tau_2^3}\int_{123} G_{12} \p_\m G_{12} G_{23} \bar\p_\m G_{23} G_{13} \eea
as given in figure 12. 

\begin{figure}[ht]
\begin{center}
\[
\mbox{\begin{picture}(250,90)(0,0)
\includegraphics[scale=.65]{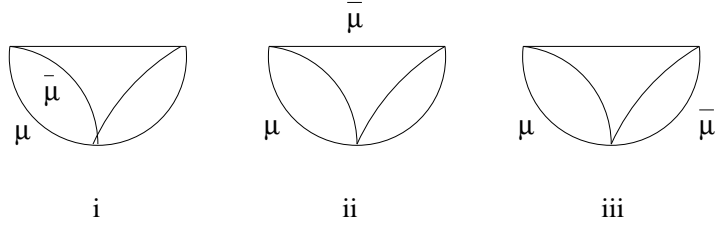}
\end{picture}}
\]
\caption{The diagrams (i) $F_{15}$, (ii) $F_{16}$ and (iii) $F_{17}$}
\end{center}
\end{figure}

We now manipulate each of these diagrams, 
leading to
\bea \label{E}F_{15} &=& E_2 E_3 - D_{1,2,2} -\frac{1}{\pi} F_{18} +\frac{1}{\pi} (F_{19} +c.c.), \non \\ F_{16} +c.c. &=& D_{1,2,2} + 2 D_{1,1,1,2} - \frac{1}{\pi} (F_{19} + c.c.), \non  \\ F_{17} &=& -\frac{3}{4} D_{1,2,2} - D_{1,1,1,2} + E_5 +\frac{1}{\pi} F_{18} +\frac{1}{2\pi} (F_{19} +c.c.)+\frac{1}{\pi} (F_{20}+c.c.),\eea
where only $F_{18}$, $F_{19}$ and $F_{20}$ involve diagrams that involve two derivatives. They are given by
\bea F_{18} &=& \frac{1}{\tau_2^3}\int_{1234} G_{12} \p_3 G_{23} G_{13}^2 \bar\p_3 G_{34} G_{14}, \non \\ F_{19} &=& \frac{1}{\tau_2^3}\int_{1234} G_{12} G_{23}^2 G_{34} \p_1 G_{14} \bar\p_2 G_{24}, \non \\ F_{20} &=& \frac{1}{\tau_2^4}\int_{12345} G_{12} G_{23} G_{34} \bar\p_5 G_{45} \p_5 G_{15} G_{25} \eea
as given in figure 13.

\begin{figure}[ht]
\begin{center}
\[
\mbox{\begin{picture}(280,80)(0,0)
\includegraphics[scale=.7]{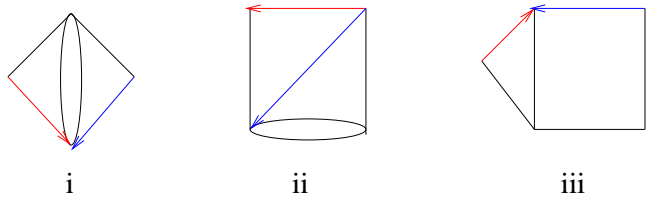}
\end{picture}}
\]
\caption{The diagrams (i) $F_{18}$, (ii) $F_{19}$ and (iii) $F_{20}$}
\end{center}
\end{figure}

We should mention that to obtain the expression for $F_{17}$ in \C{E}, we find it convenient to start from the diagram for $F_{21}$ instead given in figure 14. While this yields $\pi F_{17}$ trivially on using \C{eigen}, manipulating it by moving the derivatives along the various links appropriately leads to the expression in \C{E}.

\begin{figure}[ht]
\begin{center}
\[
\mbox{\begin{picture}(110,60)(0,0)
\includegraphics[scale=.75]{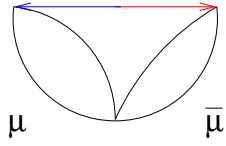}
\end{picture}}
\]
\caption{The diagram $F_{21}$}
\end{center}
\end{figure}

Thus adding the various contributions we have that
\be \Delta D_{1,2,2} = -6 D_{1,2,2} + 8 E_5 + 4 E_2 E_3 + \frac{4}{\pi} F_{18} + \frac{4}{\pi}(F_{19} + c.c.) + \frac{8}{\pi}(F_{20} +c.c.).\ee

Now let us consider the diagram $F_{18}$. To evaluate it, we start with the diagram $F_{22}$ instead which is defined by
\be F_{22} = \frac{1}{\tau_2^3} \int_{12345} \p_1 G_{12} G_{24} G_{45} \bar\p_1 G_{15} G_{13} \p_1 G_{13} \bar\p_4 G_{34} \ee
as given in figure 15. Again this trivially gives us 
\be F_{22} = \frac{\pi}{2} F_{18} - \frac{\pi^2}{2} E_2 E_3.\ee

\begin{figure}[ht]
\begin{center}
\[
\mbox{\begin{picture}(200,110)(0,0)
\includegraphics[scale=.95]{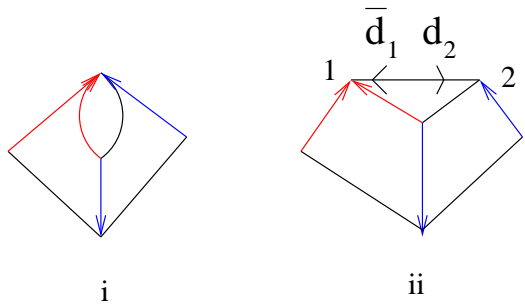}
\end{picture}}
\]
\caption{The diagrams (i) $F_{22}$ and (ii) $F_{23}$}
\end{center}
\end{figure}

Next we calculate $F_{22}$ differently. To do so, we find it convenient to start with the diagram $F_{23}$ defined by
\be F_{23} = \frac{1}{\tau_2^3}\int_{123456} \bar\p_1 \p_2 G_{12} \p_1 G_{13} G_{35} G_{56} \bar\p_2 G_{26} \p_1 G_{14} G_{24} \bar\p_5 G_{45}\ee
as given in figure 15. 
Evaluating $F_{23}$ by using \C{eigen} for the derivatives on the same link we trivially get that
\be F_{23} = \pi F_{22} + \pi^3 E_5 -\pi^3 D_{1,1,1,1;1} +\pi^2 F_{20}^*. \ee
Also evaluating it by passing the derivatives through the various links appropriately gives us that
\be F_{23} = \frac{\pi^3}{2} D_{1,2,2}+2\pi^2 F_{20}^* +\pi^2 F_{20} -\frac{\pi^2}{2} (F_{19} +c.c.).\ee
Substituting the various expressions, we obtain 
\be \label{18}F_{18} = \pi E_2 E_3 +2 \pi D_{1,1,1,1;1} -2 \pi E_5 +\pi D_{1,2,2} -(F_{19}+c.c.)+2(F_{20}+c.c.).\ee

This gives us the equation
\be \Delta D_{1,2,2} = -2 D_{1,2,2} + 8 D_{1,1,1,1;1} + 8 E_2 E_3 +\frac{16}{\pi}(F_{20}+c.c.)\ee
where $F_{20}$ is the only term that has two derivatives.

However that can be simplified further using the relation
\be \label{20}\frac{1}{\pi}(F_{20} +c.c.) = D_{1,1,1,1;1} +D_{1,1,1,2} + E_5 - E_2 E_3\ee
to get the Poisson equation
\be \label{R3}(\Delta +2)D_{1,2,2} = 24 D_{1,1,1,1;1} +16 D_{1,1,1,2} + 16 E_5 - 8 E_2 E_3\ee
involving only modular graph functions with no derivatives. 

Thus using \C{R1}, \C{R2} and \C{R3} and the asymptotic expansions, we get the relation
\be \label{x1}10 D_{1,2,2} = 20 D_{1,1,1,2} -4 E_5 + 3 \zeta(5).\ee
as conjectured in~\cite{D'Hoker:2015foa}.

Note that \C{R3} can also be written as
\be (\Delta -6)D_{1,2,2} = \frac{144}{5} E_5 -8 E_2 E_3 -\frac{8}{5} \zeta(5)\ee
which is the original form of the conjectured equation in~\cite{D'Hoker:2015foa}.

\subsection{The Poisson equation for $D_{1,1,3}$}

We next obtain the Poisson equation for $D_{1,1,3}$. To start with, we consider the diagrams $F_{24}$, $F_{25}$ and $F_{26}$ given in figure 16 which arise in the Poisson equation for $D_{1,1,3}$. Clearly $F_{24}$ and $F_{25}$ arise when $D_{1,1,3}$ is varied using \C{onevar}. We shall see the role $F_{26}$ plays in the Poisson equation later.

\begin{figure}[ht]
\begin{center}
\[
\mbox{\begin{picture}(250,90)(0,0)
\includegraphics[scale=.7]{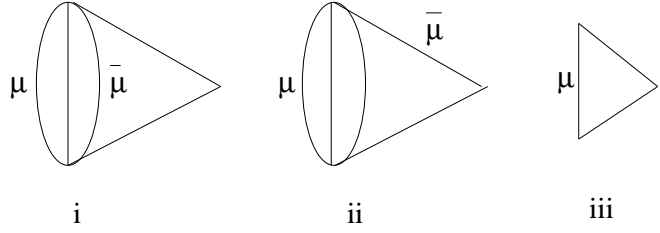}
\end{picture}}
\]
\caption{The diagrams (i) $F_{24}$, (ii) $F_{25}$ and (iii) $F_{26}$}
\end{center}
\end{figure}

To obtain the Poisson equation for $D_{1,1,3}$, we first consider the diagram $F_{27}$ given by
\be F_{27} = \frac{1}{\tau_2^3} \int_{1234} G_{13} \p_\m G_{13} \bar\p_1 \p_2 G_{12} G_{24} G_{34} \bar\p_\m G_{23}. \ee  
Again evaluating it trivially using \C{eigen}, we get that
\be F_{27} = \pi F_{24} - \pi F_6 F_{26}^*,\ee
while evaluating it by moving the derivatives along the links we get that
\bea \label{R4}\frac{1}{\pi}F_{27} &=&  F_6^* F_{26} +\frac{1}{2} D_{1,1,1,1;1} + 2 D_{1,1,3} - 2E_2 E_3 - 2 E_2 D_3 + 2D_{1,1,1,2} \non \\ &&- \frac{2}{\pi} F_{18} +\frac{1}{\pi}(F_{19} + c.c.)-\frac{2}{\pi}(F_{20} +c.c.).\eea

\begin{figure}[ht]
\begin{center}
\[
\mbox{\begin{picture}(250,120)(0,0)
\includegraphics[scale=.65]{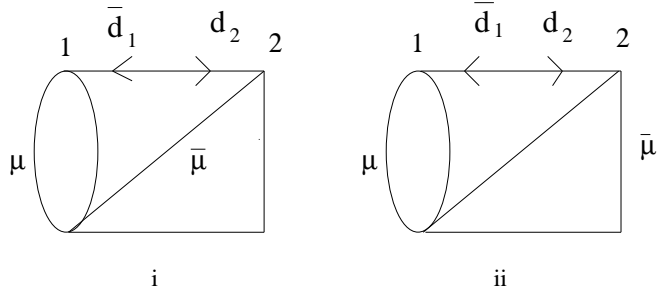}
\end{picture}}
\]
\caption{The diagrams (i) $F_{27}$ and (ii) $F_{30}$}
\end{center}
\end{figure}

In obtaining \C{R4} at an intermediate step, it is necessary to evaluate figure $F_{28}$ defined by
\be F_{28} = \frac{1}{\tau_2^3} \int_{12345} \p_1 G_{12} G_{25} \p_1 G_{13} G_{35} \bar\p_1 G_{15} \bar\p_1 G_{14} G_{45} \ee
as given in figure 18. 

\begin{figure}[ht]
\begin{center}
\[
\mbox{\begin{picture}(140,80)(0,0)
\includegraphics[scale=.6]{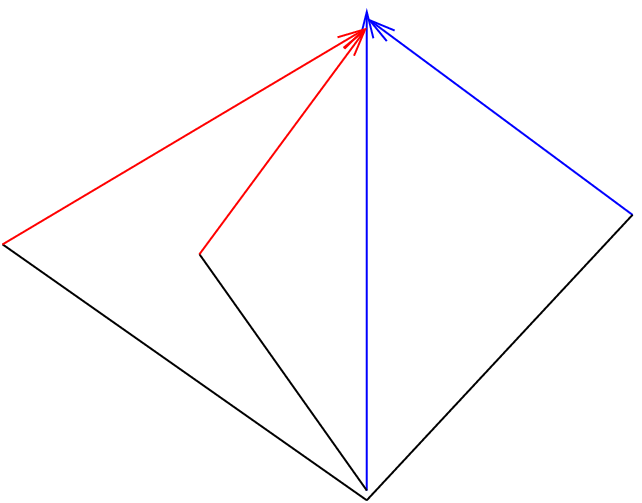}
\end{picture}}
\]
\caption{The diagram $F_{28}$}
\end{center}
\end{figure}

To do so, we start with the figure 29 instead, defined by
\be F_{29} = \frac{1}{\tau_2^3}\int_{123456} \p_1 G_{12} G_{23} \p_1 G_{16} G_{36} \bar\p_5 G_{35} \bar\p_5 G_{45} G_{34} \bar\p_1 \p_5 G_{15}\ee
as given in figure 19.

\begin{figure}[ht]
\begin{center}
\[
\mbox{\begin{picture}(140,90)(0,0)
\includegraphics[scale=.7]{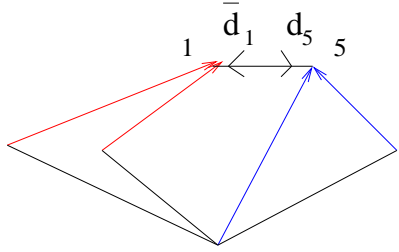}
\end{picture}}
\]
\caption{The diagram $F_{29}$}
\end{center}
\end{figure}

Using \C{eigen} and evaluating it trivially gives
\be F_{29} = \pi F_{28} -\pi^3 F_6^* F_{26},\ee
while it also gives
\be F_{29} = -2\pi^2 F_{20} -\pi^2 F_{19} +\pi^3 D_{1,2,2}\ee
on moving the derivatives along the various links.

Thus using \C{R4} we get that
\bea \label{f1}F_{24} - (F_6 F_{26}^* + c.c.) &=& \frac{1}{2} D_{1,1,1,1;1} + 2 D_{1,1,3} - 2E_2 E_3 - 2 E_2 D_3 + 2D_{1,1,1,2} \non \\ &&- \frac{2}{\pi} F_{18} +\frac{1}{\pi}(F_{19} + c.c.)-\frac{2}{\pi}(F_{20} +c.c.).\eea

We next calculate $F_{25}$. To do so, we find it convenient to consider figure $F_{30}$ defined by
\be F_{30} = \frac{1}{\tau_2^3} \int_{1234} G_{13} \p_\m G_{13} \bar\p_1 \p_2 G_{12} \bar\p_\m G_{24} G_{34}  G_{23} \ee
as given in figure 17. 

Using \C{eigen} it trivially yields
\be \label{r5}F_{30} = \pi F_{25} - \pi F_6 F_{26}^*,\ee
while evaluating it by moving the derivatives along the links gives us
\bea \label{R5}\frac{1}{\pi} F_{30} &=& F_6 F_{26}^* -\frac{1}{6} D_{1,1,3} -\frac{1}{2} E_2 E_3 +\frac{1}{2} E_2 D_3 -\frac{3}{2} D_{1,1,1,2} - \frac{1}{2} D_{1,2,2} \non \\ &&-2 D_{1,1,1,1;1} + 2 E_5 +\frac{1}{\pi} F_{18} -\frac{1}{2\pi} (F_{19} - c.c.) + \frac{2}{\pi}(F_{20} + c.c.).\eea
In obtaining \C{R5}, at an intermediate step we evaluated the diagram $F_{31}$ defined by
\be F_{31} = \frac{1}{\tau_2^3}\int_{12345} G_{12} G_{13} \p_4 G_{34} \p_4 G_{14} \bar\p_4 G_{14} G_{25} \bar\p_4 G_{45}\ee
as given in figure 20.

\begin{figure}[ht]
\begin{center}
\[
\mbox{\begin{picture}(120,70)(0,0)
\includegraphics[scale=.8]{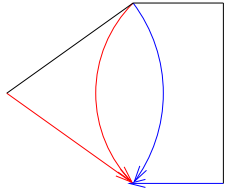}
\end{picture}}
\]
\caption{The diagram $F_{31}$}
\end{center}
\end{figure}

To do so, we start with diagram $F_{32}$ instead defined by
\be F_{32} = \frac{1}{\tau_2^3} \int_{123456} G_{12} G_{13} \p_4 G_{34} \p_4 G_{14} \bar\p_5 G_{15} G_{26} \bar\p_5 G_{56} \bar\p_4 \p_5 G_{45}\ee
as given in figure 21.

\begin{figure}[ht]
\begin{center}
\[
\mbox{\begin{picture}(120,100)(0,0)
\includegraphics[scale=.7]{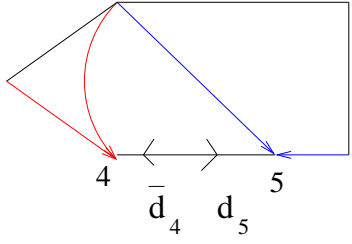}
\end{picture}}
\]
\caption{The diagram $F_{32}$}
\end{center}
\end{figure}

Using \C{eigen} and evaluating it trivially gives us
\be \label{31}F_{32} = \pi F_{31} - \pi^3 F_6 F_{26}^*,\ee
while it also gives us
\be \label{32}\frac{1}{\pi^3}F_{32} = - D_{1,1,1,2} + E_5 + \frac{1}{2} E_2 D_3 - D_{1,1,1,1;1} -\frac{1}{2\pi} F_{19}+\frac{1}{\pi} F_{20}^*\ee
on moving the derivatives along the various links.

Thus from \C{r5} and \C{R5} we get that
\bea \label{f2}F_{25} -2 F_6 F_{26}^* + c.c. &=& -\frac{1}{3} D_{1,1,3} - 3 D_{1,1,1,2} - D_{1,2,2} - 4 D_{1,1,1,1;1} + 4 E_5 \non \\ &&- E_2 E_3 + E_2 D_3  +\frac{2}{\pi} F_{18} +\frac{4}{\pi}(F_{20} +c.c.).\eea

Now
\bea \label{M}&&\Delta (D_{1,1,3} - 3 E_2 E_3) = \p_\m \bar\p_\m (D_{1,1,3} - 3 E_2 E_3) \non \\ &&= 6\Big[ F_{24} - (F_6 F_{26}^* + c.c.)\Big] + 6(F_{25} - 2 F_6 F_{26}^*+c.c.) + 2 D_{1,1,3} - 24 E_2 E_3\eea
where we have chosen the relative coefficients between $D_{1,1,3}$ and $E_2 E_3$ such that the action of $\Delta$ on it precisely involves the combinations $F_{24} - (F_6 F_{26}^* + c.c.)$ and $F_{25} - 2 F_6 F_{26}^* + c.c.$ which are given by \C{f1} and \C{f2} respectively.

We now simplify this equation by using \C{f1} and \C{f2}. Among the terms involving two derivatives in the last line of \C{M}, the contribution from $F_{18}$ cancels, while for $F_{20}+c.c.$ we use the relation \C{20}. Also for $F_{19}+c.c.$  we use the relation
\be \label{19}\frac{1}{\pi}(F_{19}+c.c.) = E_2 E_3 + E_2 D_3 - D_{1,1,1,2} + D_{1,2,2}- D_{1,1,3}.\ee 
Finally we get the Poisson equation
\be \label{113}(\Delta -6)(D_{1,1,3} - 3 E_2 E_3) = 36 E_5 - 9 D_{1,1,1,1;1}   - 30 E_2 E_3\ee
for the diagram $D_{1,1,3}$. 
Thus from the above equation, \C{R1} and \C{R2} and the asymptotic expansions, we have that
\be \label{x2} 40 D_{1,1,3} = 300 D_{1,1,1,2} + 120 E_2 E_3 -276 E_5 + 7\zeta(5)\ee
as conjectured in~\cite{D'Hoker:2015foa}. We can also write \C{113} as
\be (\Delta -6)(D_{1,1,3} - 3 E_2 E_3) = \frac{162}{5} E_5 - 30 E_2 E_3 - \frac{3}{10}\zeta(5).\ee

\subsection{The Poisson equation for $D_{5}$}

Finally we obtain the Poisson equation for $D_5$. We list the diagrams $F_{33}$ and $F_{34}$ given in figure 22, which will be relevant for our purposes. While $F_{33}$ arises simply on varying $D_5$ using \C{onevar}, we shall see that $F_{34}$ also arises in the Poisson equation.

\begin{figure}[ht]
\begin{center}
\[
\mbox{\begin{picture}(180,110)(0,0)
\includegraphics[scale=.7]{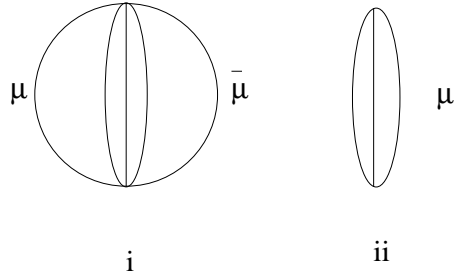}
\end{picture}}
\]
\caption{The diagrams (i) $F_{33}$ and (ii) $F_{34}$}
\end{center}
\end{figure}

We have that
\be \pi^2 F_{33} = 3 F_{35}+\frac{\pi^2}{4}D_5,\ee
where the diagram $F_{35}$ is defined by
\be F_{35} = \frac{1}{\tau_2^2} \int_{1234} \p_2 G_{12} G_{23} G_{13}^2 \p_3 G_{13} \bar\p_4 G_{14} \bar\p_4 G_{34}\ee
as given in figure 23.

\begin{figure}[ht]
\begin{center}
\[
\mbox{\begin{picture}(120,70)(0,0)
\includegraphics[scale=.8]{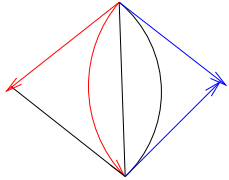}
\end{picture}}
\]
\caption{The diagram $F_{35}$}
\end{center}
\end{figure}

To evaluate $F_{35}$ we find it convenient to start with the diagram $F_{36}$ defined by
\be F_{36} = \frac{1}{\tau_2^2} \int_{12345} \p_2 G_{12} G_{23} \p_3 G_{13} G_{35}^2 \bar\p_4 G_{34} \bar\p_4 G_{45} \bar\p_1 \p_5 G_{15}\ee
as given in figure 24.

\begin{figure}[ht]
\begin{center}
\[
\mbox{\begin{picture}(120,90)(0,0)
\includegraphics[scale=.65]{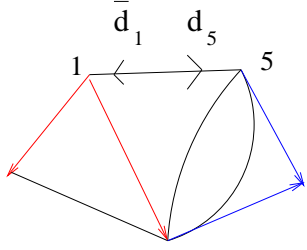}
\end{picture}}
\]
\caption{The diagram $F_{36}$}
\end{center}
\end{figure}

On using \C{eigen}, this trivially yields
\be F_{36} = \pi F_{35} - \pi^3 F_6 F_{34}^*,\ee
while it also yields
\be \frac{1}{\pi^3}F_{36}= \frac{1}{12} D_5 - \frac{1}{3} E_2 D_3 - D_{1,2,2} - \frac{2}{3} D_{1,1,3} +\frac{1}{\pi} (F_{18} + F_{19}^* +2 F_{20}^*) +\frac{1}{\pi^2}(2 F_{31}^* -  F_{37})\ee
on moving the derivatives through the links. Here the diagram  $F_{37}$ is defined by
\be F_{37} = \frac{1}{\tau_2^2} \int_{1234} \p_2 G_{12} \p_2 G_{24} \bar\p_2 G_{24} \bar\p_2 G_{23} G_{34} G_{14}^2\ee
as given in figure 25.

\begin{figure}[ht]
\begin{center}
\[
\mbox{\begin{picture}(100,60)(0,0)
\includegraphics[scale=.75]{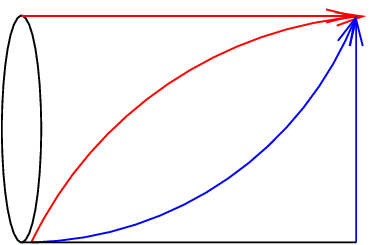}
\end{picture}}
\]
\caption{The diagram $F_{37}$}
\end{center}
\end{figure}

To evaluate $F_{37}$ we start with the diagram $F_{38}$ instead defined by
\be F_{38} = \frac{1}{\tau_2^2} \int_{12345} G_{15}^2 \p_2 G_{12} \p_2 G_{25} \bar\p_3 G_{35} \bar\p_3 G_{34} G_{45} \bar\p_2 \p_3 G_{23}\ee
as given in figure 26.

\begin{figure}[ht]
\begin{center}
\[
\mbox{\begin{picture}(100,90)(0,0)
\includegraphics[scale=.65]{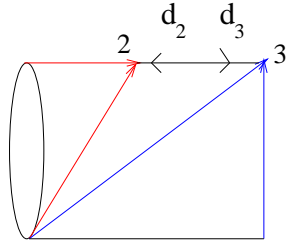}
\end{picture}}
\]
\caption{The diagram $F_{38}$}
\end{center}
\end{figure}

Using \C{eigen} this trivially evaluates to
\be F_{38} = \pi F_{37} -\pi^3 F_6^* F_{34},\ee
while it also yields
\be \frac{1}{\pi^3} F_{38} = \frac{1}{6} D_5 -\frac{2}{3} E_2 D_3 - \frac{1}{3} D_{1,1,3} + E_2 E_3 - \frac{1}{2} D_{1,2,2} + D_{1,1,1,2}\ee
on moving the derivatives through the various links.

Thus substituting the various expressions, we get that
\bea \frac{1}{3} F_{33} - (F_6 F_{34}^* + c.c.) &=& \frac{4}{3} E_2 D_3 -\frac{1}{2} D_{1,2,2} - \frac{1}{3} D_{1,1,3} -E_2 E_3 - 3 D_{1,1,1,2} \non \\ &&+ 2 E_5 -2 D_{1,1,1,1;1}+\frac{1}{\pi} F_{18}+\frac{2}{\pi}(F_{20} + c.c.).\eea
where we have substituted the expression for $F_{31}$ obtained from \C{31} and \C{32}. We have also used $\p_\m D_3 = \p_\m E_3$ which follows trivially from \C{1}. 

Finally using \C{18}, \C{20} and \C{19} we get that
\bea F_{33} - 3 (F_6 F_{34}^* +c.c.) &=& E_2 D_3 -\frac{3}{2} D_{1,2,2} + 2 D_{1,1,3} -15 E_2 E_3 + 6 D_{1,1,1,2} \non \\ &&+ 12 (E_5 + D_{1,1,1,1;1}).\eea
 
Thus we obtain the Poisson equation
\bea \Delta (D_5 - 10 E_2 D_3) &=& \p_\m \bar\p_\m (D_5 - 10 E_2 D_3) \non \\ &=& 20[F_{33} - 3 (F_6 F_{34}^* +c.c.)] - 20 E_2 D_3 - 60 E_2 E_3 \non \\ &=& -30 D_{1,2,2} + 40 D_{1,1,3} +120 D_{1,1,1,2} + 240(E_5 + D_{1,1,1,1;1})-360 E_2 E_3.\non \\\eea
Once again we have chosen the relative coefficient between $D_5$ and $E_2 D_3$ such that the action of $\Delta$ produces $F_{33} - 3(F_6F_{34}^*+c.c.)$ which we have determined separately.

Thus using \C{R1}, \C{x1} and \C{x2} we get the desired Poisson equation
\be \label{5}\Delta (D_5 - 10 E_2 D_3) = 360 D_{1,1,1,2} + 72 E_5 -240 E_2 E_3 + 6\zeta(5)\ee
which using \C{R2} yields
\be \Delta(D_5 -60 D_{1,1,1,2} + 48 E_5 - 10 E_2 D_3)=0,\ee
and hence
\be D_5 =60 D_{1,1,1,2} - 48 E_5 + 10 E_2 D_3 + 16 \zeta(5)\ee
using the asymptotic expansions. This relation between modular graph functions had been conjectured in~\cite{D'Hoker:2015foa}.

In fact, \C{5} can also be written as
\be (\Delta -6)(D_5 - 10 E_2 D_3)= 360 E_5 - 240 E_2 E_3 - 90 \zeta(5)\ee
which was the form of the equation originally conjectured in~\cite{D'Hoker:2015foa}.

Thus we have obtained several relations between modular graph functions with five links. Note that they also involve $E_5$, which arises in the five graviton amplitude. 

\section{Poisson equations for some diagrams with six links}

We next consider Poisson equations for certain diagrams with six links having four and five vertices. These arise in the low energy expansion of the four and five graviton amplitudes at orders $D^{12} \mathcal{R}^4$ and $D^{10}\mathcal{R}^5$ respectively. In either case, they are not the only modular graph functions that arise for these amplitudes at these orders in the derivative expansion. However, we shall see that the Poisson equations for these diagrams provide enough information for us to obtain a non--trivial relation among graphs with six links. The relevant diagrams are given in figure 27.        
\begin{figure}[ht]
\begin{center}
\[
\mbox{\begin{picture}(280,100)(0,0)
\includegraphics[scale=.65]{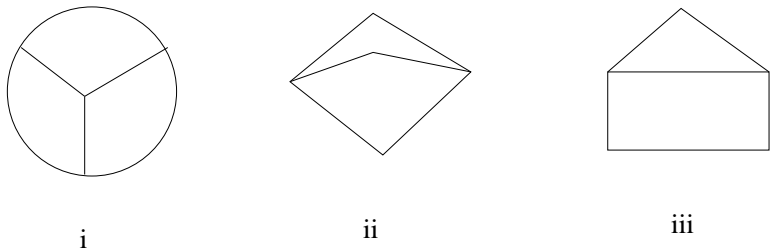}
\end{picture}}
\]
\caption{The diagrams (i) $D_{1,1,1;1,1,1}$, (ii) $C_{2,2,2}$ and (iii) $C_{1,2,3}$}
\end{center}
\end{figure}

They are given by the expressions
\bea && D_{1,1,1;1,1,1} = \frac{1}{\tau_2^4} \int_{1234} G_{12} G_{23} G_{13} G_{14} G_{24} G_{34}, \quad C_{2,2,2} = \frac{1}{\tau_2^5}\int_{12345} G_{12} G_{23} G_{34} G_{14} G_{25} G_{45}, \non \\ && C_{1,2,3} = \frac{1}{\tau_2^5} \int_{12345} G_{12} G_{23} G_{13} G_{24} G_{45} G_{35}.\eea
For the five point graphs, we use the terminology of~\cite{D'Hoker:2015foa}.

\subsection{The Poisson equation for $D_{1,1,1;1,1,1}$}

The Poisson equation for the Mercedes graph $D_{1,1,1;1,1,1}$ has been obtained in~\cite{Basu:2015ayg} and is given by
\be \label{61}(\Delta +6) D_{1,1,1;1,1,1} = 48 C_{1,2,3} + 12(E_6 -E_3^2). \ee 

\subsection{The Poisson equation for $C_{2,2,2}$}

We now obtain the Poisson equation for $C_{2,2,2}$. Proceeding as earlier, we get that
\be \Delta C_{2,2,2} = 6 F_{39} + 24 F_{40},\ee
where
\bea F_{39} &=& \frac{1}{\tau_2^5}\int_{12345} \bar\p_\mu G_{12} G_{23} G_{34} \p_\mu G_{14} G_{25} G_{45}, \non \\ F_{40} &=& \frac{1}{\tau_2^5}\int_{12345} G_{12} G_{23} \bar\p_\mu G_{34} \p_\mu G_{14} G_{25} G_{45}\eea
as given in figure 28.

\begin{figure}[ht]
\begin{center}
\[
\mbox{\begin{picture}(180,100)(0,0)
\includegraphics[scale=.65]{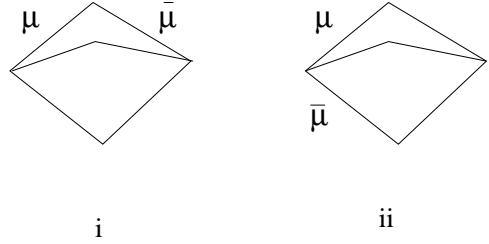}
\end{picture}}
\]
\caption{The diagrams (i) $D_{1,1,1;1,1,1}$, (ii) $C_{2,2,2}$ and (iii) $C_{1,2,3}$}
\end{center}
\end{figure}

Now using \C{onevar} and \C{eigen} we get that
\bea F_{39} &=& C_{2,2,2} , \non \\ F_{40} &=&  -C_{1,2,3} + \frac{1}{2}(E_3^2- E_6)\eea
leading to
\be \label{62}(\Delta -6) C_{2,2,2} = -24 C_{1,2,3} + 12 (E_3^2 - E_6)\ee
as obtained in~\cite{D'Hoker:2015foa} using other techniques.

\subsection{The Poisson equation for $C_{1,2,3}$}

Next we obtain the Poisson equation for $C_{1,2,3}$. Like before, we have that
\be \Delta C_{1,2,3} = 2 F_{41} + 6 F_{42} + 2 (F_{43} +c.c.) + 3 (F_{44} + c.c.) + 6(F_{45}+c.c.),\ee
where
\bea F_{41} &=& \frac{1}{\tau_2^5} \int_{12345} \p_\m G_{12} G_{23} \bar\p_\mu G_{13} G_{24} G_{45} G_{35}, \non \\ F_{42}& =& \frac{1}{\tau_2^5} \int_{12345} G_{12} G_{23} G_{13} \p_\mu G_{24} G_{45} \bar\p_\mu G_{35} , \non \\  F_{43}& =& \frac{1}{\tau_2^5} \int_{12345} \p_\mu G_{12} \bar\p_\mu G_{23} G_{13} G_{24} G_{45} G_{35} , \non \\ F_{44} &=& \frac{1}{\tau_2^5} \int_{12345} G_{12} \p_\mu G_{23} G_{13} G_{24} \bar\p_\mu G_{45} G_{35} , \non \\ F_{45} &=& \frac{1}{\tau_2^5} \int_{12345} \p_\mu G_{12} G_{23} G_{13} \bar\p_\mu G_{24} G_{45} G_{35} \eea
as given in figure 29. 

\begin{figure}[ht]
\begin{center}
\[
\mbox{\begin{picture}(360,100)(0,0)
\includegraphics[scale=.6]{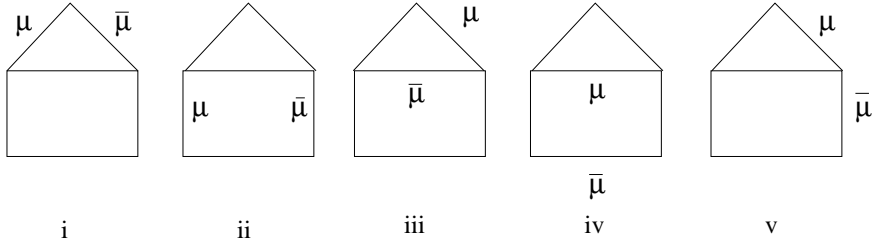}
\end{picture}}
\]
\caption{The diagrams (i) $F_{41}$, (ii) $F_{42}$, (iii) $F_{43}$, (iv) $F_{44}$ and (v) $F_{45}$}
\end{center}
\end{figure}

Again using \C{onevar} and \C{eigen}, we get that
\bea F_{41} &=& F_{42} = C_{1,2,3}, \non \\ F_{43} &=& - C_{2,2,2} +\frac{1}{2}(E_3^2 - E_6), \non \\ F_{44} + 2 F_{45} &=& \frac{1}{2} C_{2,2,2} + 3 E_6 - E_3^3, \eea
giving us
\be \label{63} (\Delta -8) C_{1,2,3} = - C_{2,2,2} -4 (E_3^2 - 4E_6)\ee
as obtained in~\cite{D'Hoker:2015foa} using other techniques.

Thus from \C{61}, \C{62} and \C{63} we have that 
\be (\Delta +6) (3 D_{1,1,1;1,1,1} -12 C_{1,2,3} - C_{2,2,2} +4 E_6)=0\ee
leading to the relation among the modular graph functions with six links\footnote{The relation also involves $E_6$ which arises in the six graviton amplitude.}
\be 3 D_{1,1,1;1,1,1} =12 C_{1,2,3} + C_{2,2,2} -4 E_6\ee
on using the asymptotic expansions.

Thus our analysis proves various Poisson equations satisfied by the modular graph functions, which also lead to several relations among them. We have used various auxiliary graphs to simplify our calculations, and also simplified several expressions by moving the derivatives through the links appropriately. Clearly this procedure is not unique and one can proceed in different ways to obtain the relations. It would be interesting to generalize the analysis to higher orders in the derivative expansion and also for modular graph functions involving derivatives of Green functions. It would also be useful to obtain relations involving higher genus string amplitudes.    

\section{Discussion}

Our analysis of obtaining Poisson equations for modular graph functions fits into the general scheme of calculating perturbative string amplitudes, and generalize various features of the amplitudes at tree level. Low momentum expansion of the tree level amplitudes yields various multi--zeta values on performing the integrals over the insertion points of the integrated vertex operators. At a fixed order in the momentum expansion, this structure simplifies on using various identities between the multi--zeta values, which follows from a detailed analysis of the number of basis elements of fixed transcendentality. Thus it is essential to understand the basis elements in detail to analyze tree level amplitudes. 

Clearly it is important to generalize this structure at higher loops in string perturbation theory, which is something this work addresses. At a fixed order in the momentum expansion of the one loop amplitude, we see that there are several topologically distinct graphs given by the various ways the Green functions connect the vertices. However the relations among the modular graph functions show that they are not all independent and the number of basis elements is far less than the number of topologically distinct modular graph functions. This analysis, as well as its generalization at higher loops, is crucial in simplifying the structure of the integrands of the loop amplitudes, which can then be integrated over the moduli space on using Poisson equations. It is by itself interesting mathematically to obtain relations between graphs at various loops.               

In the Poisson equations for the various graphs, we see that the number of links is conserved while the number of vertices are not. The source terms involve graphs that have been obtained at lower orders in the momentum expansion. This feature must survive to all orders, hence allowing us to recursively solve the equations. Note that we have considered graphs which only involve the scalar Green function as links, and not their derivatives. Such graphs that involve derivatives arise in the low momentum expansion of higher point multi--graviton amplitudes. It is interesting to see if such graphs can be expressed in terms of those which only involve scalar Green functions and not their derivatives. If this is not always the case, they yield new elements in the basis of modular graph functions, analyzing which is essential to get a complelete basis of independent graphs. Generalizing this analysis to higher loops is crucial to obtain results in perturbative string theory. 

Finally, the integrands of the amplitudes we have considered simplify because of maximal supersymmetry. In order to calculate string amplitudes with lesser supersymmetry, one has to generalize the techniques we have discussed to obtain Poisson equations for the various graphs. It would be important to analyze the basis elements that arise in such cases. Clearly, the analysis gets more involved as the amount of supersymmetry is reduced. 

Thus we see that for both calculating perturbative string amplitudes as well as from the point of view of mathematics, it is important to obtain relations between modular graphs at various loops and very little is understood. Hence a better understanding of this structure is desirable.


\begin{thebibliography}{10}

\bibitem{Green:1999pv}
M.~B. Green and P.~Vanhove, ``{The Low-energy expansion of the one loop type II
  superstring amplitude},'' {\em Phys.Rev.} {\bf D61} (2000) 104011,
\href{http://www.arXiv.org/abs/hep-th/9910056}{{\tt hep-th/9910056}}.

\bibitem{Green:2013bza}
M.~B. Green, C.~R. Mafra, and O.~Schlotterer, ``{Multiparticle one-loop
  amplitudes and S-duality in closed superstring theory},'' {\em JHEP} {\bf 10}
  (2013) 188,
\href{http://www.arXiv.org/abs/1307.3534}{{\tt 1307.3534}}.

\bibitem{D'Hoker:2015foa}
E.~D'Hoker, M.~B. Green, and P.~Vanhove, ``{On the modular structure of the
  genus-one Type II superstring low energy expansion},'' {\em JHEP} {\bf 08}
  (2015) 041,
\href{http://www.arXiv.org/abs/1502.06698}{{\tt 1502.06698}}.

\bibitem{D'Hoker:2015zfa}
E.~D'Hoker, M.~B. Green, and P.~Vanhove, ``{Proof of a modular relation between
  1-, 2- and 3-loop Feynman diagrams on a torus},''
\href{http://www.arXiv.org/abs/1509.00363}{{\tt 1509.00363}}.

\bibitem{D'Hoker:2016jac}
E.~D'Hoker and M.~B. Green, ``{Identities between Modular Graph Forms},''
\href{http://www.arXiv.org/abs/1603.00839}{{\tt 1603.00839}}.

\bibitem{D'Hoker:2015qmf}
E.~D'Hoker, M.~B. Green, O.~Gurdogan, and P.~Vanhove, ``{Modular Graph
  Functions},''
\href{http://www.arXiv.org/abs/1512.06779}{{\tt 1512.06779}}.

\bibitem{Basu:2015ayg}
A.~Basu, ``{Poisson equation for the Mercedes diagram in string theory at genus
  one},'' {\em Class. Quant. Grav.} {\bf 33} (2016), no.~5, 055005,
\href{http://www.arXiv.org/abs/1511.07455}{{\tt 1511.07455}}.

\bibitem{Basu:2016xrt}
A.~Basu, ``{Poisson equation for the three loop ladder diagram in string theory
  at genus one},''
\href{http://www.arXiv.org/abs/1606.02203}{{\tt 1606.02203}}.

\bibitem{Green:2008uj}
M.~B. Green, J.~G. Russo, and P.~Vanhove, ``{Low energy expansion of the
  four-particle genus-one amplitude in type II superstring theory},'' {\em
  JHEP} {\bf 0802} (2008) 020,
\href{http://www.arXiv.org/abs/0801.0322}{{\tt 0801.0322}}.

\bibitem{Basu:2016fpd}
A.~Basu, ``{Non-BPS interactions from the type II one loop four graviton
  amplitude},'' {\em Class. Quant. Grav.} {\bf 33} (2016), no.~12, 125028,
\href{http://www.arXiv.org/abs/1601.04260}{{\tt 1601.04260}}.

\end{thebibliography}

\providecommand{\href}[2]{#2}\begingroup\raggedright\endgroup

\end{document}